\documentclass[onecolumn, fleqn,usenatbib]{mnras}
\usepackage[T1]{fontenc}
\usepackage{ae,aecompl}

\usepackage{graphicx}	% Including figure files
\usepackage{amsmath}	% Advanced maths commands
\usepackage{amssymb}	% Extra maths symbols
\usepackage{color}
\definecolor{orange}{rgb}{1.0,0.5,0.0}

\title[CiC in IllustrisTNG]{Counts-in-Cells of subhaloes in the IllustrisTNG simulations: the role of baryonic physics}

\author[Dantas, C. C.]{
Christine C. Dantas$^{1}$\thanks{E-mail: christine.dantas@inpe.br}
\\
$^{1}$Divis\~ao de Astrof\'{\i}sica (INPE-MCTI), 
S\~ao Jos\'e dos Campos, 12227-010, SP, Brazil 
}

\date{Accepted XXX. Received YYY; in original form \today}

\pubyear{2020}

\begin{document}
\label{firstpage}
\pagerange{\pageref{firstpage}--\pageref{lastpage}}
\maketitle

\begin{abstract}

We present an analysis of the Counts-in-Cells (CiC) statistics of subhaloes in the publicly available IllustrisTNG cosmological simulations (TNG100-1, TNG100-3 and TNG300-3), considering their full and dark-only versions, in redshifts ranging from $z = 0$ to $z=5$, and different cell sizes. We evaluated two CiC models: the gravitational quasi-equilibrium distribution (GQED) and the negative binomial distribution (NBD), both presenting good fits, with small detectable differences in the presence of baryons. Scaling and time dependencies of the best-fit parameters showed similar trends compared with the literature. We derived a matter density-in-cells probability distribution function (PDF), associated with the GQED, which was compared to the PDF given in \cite{Uhl16}, for the IllustrisTNG 100-3-Dark run at $z=0$. Our results indicate that the simplest gravithermodynamical assumptions of the GQED model hold in the presence of baryonic dissipation. Interestingly, the smoothed (density-in-cells) version of the GQED is also adequate for describing the dark matter one-point statistics of subhaloes and converges, to subpercentage levels (for an interval of parameters), to the Uhlemann et al. PDF in the high density range. 

\end{abstract}


%%%%%%%%%%%%%%%%%%%%%%%%%%%%%%%%%%%%%%%%%%%%%%%%%%
% 1 - INTRODUCTION
%%%%%%%%%%%%%%%%%%%%%%%%%%%%%%%%%%%%%%%%%%%%%%%%%%

\defcitealias{PLANCK16}{Planck Collaboration 2016}

\section{Introduction \label{INTRO}}

The recognition that the Universe contains large-scale structures involved a long process of discovery, from initial speculations to physical theories, guided by increasingly refined data from observational surveys \citep[see a historical account in, e.g.,][]{SaslawBOOK00}. These structures, composed of smaller gravitational units (clusters of galaxies, galaxies, etc.),  probably formed from small initial fluctuations in the early Universe, and evolved through gravitational instability in an expanding space-time. Observations such as the cosmic microwave background radiation and baryon acoustic oscillation signals \citep{PLANCK16}, high-redshift Type Ia supernovae \citep{Rie98,Per99}, etc., converged to a spatially flat $\Lambda$CDM cosmological model, which however has recently been under significant tension due to revealing inconsistencies \citep[e.g.][]{Qui19}. 

Our current understanding indicates that galaxies (and other observables such as quasars, line intensities maps, diffuse backgrounds, etc.)  are only tracers of the large-scale structures, with the mass component being dominated by dark matter (DM), evolving in an accelerated expanding background (due to some form of dark energy). In this $\Lambda$CDM cosmology, large-scale structures are formed hierarchically, in which DM haloes are assembled from the merging of smaller structures. The relation between the distribution of tracers and the underlying distribution of (total) matter, termed ``bias'' \citep{Wei04,Des18}, its nature and how it evolves in time, is of fundamental importance, not only for the understanding and characterization of large-scale structures, but also as tests for cosmological models. 

The study of large-scale structures and its nonlinear evolution also relies on statistical tools such as the power spectra and correlation functions \citep{Pee80,Ber02}. Another approach is given by the Counts-in-Cells (CiC) probability distribution function (PDF)  \citep{Efs90,Sza98,SaslawBOOK85,SaslawBOOK00}, in which discrete objects such as galaxies are counted inside cells of fixed size and shape in an ensemble. The CiC can also be expressed as an appropriately smoothed density-in-cells distribution, and it can be derived from fundamental theories of gravitational clustering and/or estimated from survey data \citep[e.g.][]{Uhl16,Uhl18,Sal19}.  

The gravitational quasi-equilibrium distribution (GQED) describes the CiC statistics of N-point masses in an expanding universe. The GQED was first derived by \cite{Sas84}, leading to several subsequent studies and refinements  (summarized in the book by \citealt{SaslawBOOK00}, following an earlier foundational exposition in \citealt{SaslawBOOK85}). In GQED, the pairwise properties of the gravitational potential are consistently introduced into a gravithermodynamical theory, under certain hypotheses, but it has also been subsequently derived from statistical mechanics principles by \cite{Ahm02}.  Another CiC PDF of interest is the negative binomial distribution (NBD), first proposed in a cosmological context by \cite{Car83}, and subsequently explored in \cite{Eli92} \citep[see also, e.g.][]{Yan11,HGil17,WKS20}. 

The Illustris The Next Generation (TNG) project\footnote{https://www.tng-project.org/} presents a prominent series cosmological simulations \citep{Nel19,Pil18,Spr18,Nel18,Nai18,Mar18}, suitable for the study of galaxy formation and evolution, following the coupled dynamics of baryons and DM through a state-of-the-art magnetohydrodynamical numerical code \citep[{\tt AREPO}, ][]{Spr10}.  Dark-only counterparts to the full runs are also available, with related studies on the large-scale clustering and bias  \citep[e.g.][]{Spr18,BaryonsWeb19,BaryonsWeb20,Dor20}.
 
In the present work, we investigated the one-point statistics of subhaloes in the publicly available IllustrisTNG simulations (full and dark-only versions of the TNG100-1, TNG100-3 and TNG300-3 runs), in terms of the CiC statistics (the GQED and the NBD), and in terms of the density-in-cells statistics. In the latter case, we derived a density-in-cell PDF for the GQED model, obtained from its CiC counterpart under simplified assumptions (hereon, denoted by ``dGQED PDF'', the ``d'' prefix referring to density-in-cells). This model was compared to the DM density-in-cells PDF proposed by Uhlemann et al. (hereon denoted ``UHL PDF''; \citealt{Uhl16,Uhl18}). 

Our paper is organized as follows: in Sec. \ref{METH}, we present our motivation and a few clarifications. This section also presents the selection of the IllustrisTNG runs, a summary of the methods used to extract the one-point statistics from the simulations, the models, and the fitting procedure. In Sec. \ref{RES}, we present our results: the CiC distributions and the related GQED and NBD best-fit parameters, as well as a comparative analysis of the obtained density-in-cells PDFs. In Sec. \ref{DIS}, we summarize and discuss our results. The Appendix provides a brief study of the residues of the GQED and NBD fits. The cosmology parameters used in the present paper are those defined in the IllustrisTNG simulations, namely: $\Omega_{\Lambda,0}= 0.6911, \Omega_{{\rm m},0} = \Omega_{{\rm DM},0} + \Omega_{{\rm b},0} = 0.3089, \Omega_{{\rm b},0} = 0.0486, \sigma_{8} = 0.8159, n_s = 0.9667$ and $h = 0.6774$ \citepalias{PLANCK16}.

%%%%%%%%%%%%%%%%%%%%%%%%%%%%%%%%%%%%%%%%%%%%%%%%%%
% 2 - METHODS
%%%%%%%%%%%%%%%%%%%%%%%%%%%%%%%%%%%%%%%%%%%%%%%%%%

\section{Motivation and Methodology \label{METH}}

%%%%%%%%%%%%%%%%%%%%%%%%%%%%%%%%%%%%%%%%%%%%%%%%%%

\subsection{Motivation\label{MOTIV}}

Are the gravithermodynamical assumptions underlying the GQED applicable to the distribution of dark matter subhaloes in the IllustrisTNG? If so, then, in the case of the full runs, does the complexity of the incorporated physical processes and gravo-magnetohydrodynamics introduce secondary corrections to the GQED? Do other models, not derived (in principle, at least) from gravithermodynamical conditions, provide similar predictions for the one-point statistics, with or without baryonic dissipation? Are additional elements necessary for describing the combined clustering of dark and baryonic matter?

This work attempts to extend the scope of applications of the IllustrisTNG simulations into tests of gravithermodynamical theory, under complex, multi-component clustering dynamics. Our investigation focuses on the simplest GQED model, ie., the one introduced by \cite{Sas84} (and subsequently explored in various directions). Before we proceed,  we would like to highlight two important points:

\begin{enumerate}
\item{{\it The GQED is not an empirical model}, but based on theoretical considerations. It can be derived consistently either from gravitational thermodynamics or from statistical mechanics \citep[eg.,][]{Ahm02}. There is a significant body of literature about the GQED, developed for more than $30$ years, including extensions, as well as numerical and observational results, which were summarized in a book \citep{SaslawBOOK00}.  The original GQED theory assumes that galaxies are N-body point masses in an infinite, expanding universe model, but this theoretical idealization has been proved to be adequate to first order in several studies. }
\item{{\it The simplest form of the GQED, as well as the NBD, have no free parameters}. The GQED  modifies the Poisson PDF due to the presence of correlations, and is described by two parameters: the clustering parameter $b$, and the average number of galaxies per cell, $\bar{N}$. The NBD is also similarly described. Clearly, $\bar{N}$ can be obtained directly from the data, but it is also the case for the clustering parameter, as we explain in the next section. However, we found that the parameters were more straightforwardly obtained from fittings, hence this was our methodological choice. In any case,  it is important to emphasize that {\it a fitting to the QGED automatically provides an implicit test of the adequacy of the thermodynamical regime presupposed in the theory} \citep[cf.][]{She96}. On the other hand, the NBD model has been argued to be unphysical \citep{Sas96}, even though it does provide a good fit to the observed CiC PDFs \citep[e.g.][]{HGil17,WKS20}. }
\end{enumerate} 

Some additional notes that motivate our investigation follows. First, it is important to understand limiting behaviours of the PDF parameters, as they encode information on how the gravitational clustering evolves in terms of spatial regions. For instance, in smaller scales, it is fundamental to search for the effects of mergers on the predictions of the CiC statistics, and on larger scales, indications of non-Poissonian initial conditions. 
Previous investigations analyzed the potential impact of certain extensions on the parameter-free nature of the simplest form of the GQED theory. An important consideration is evolution and scale dependence of the parameters \citep[eg.,][]{She96}. Therefore, one of the motivations for the present work was to analyze the CiC parameters in terms of these dependencies. 

Second, the role of the dominant dynamical component (dark matter) is of great interest, specially whether it is individually associated with galaxies through distinct haloes, or more uniformly distributed throughout space. These properties are important clues for cosmological models.
Recently, an analysis of the potential use of the GQED and other models for selecting among different cosmologies has been conducted by \cite{WKS20}.  These authors used the $648$ Mpc $h^{-1}$ box, Dark Energy Universe Simulations  \citep[DEUS,][]{alimi2012deus}, to study the CiC PDFs of DM haloes, as a function of scale and redshift, finding nuanced features which nevertheless could be used to observationally distinguish between different cosmologies.
In the present case, the IllustrisTNG simulations are all set up with the same cosmological model, therefore our analysis addressed the potential nuances of the one-point statistics of subhaloes under a fixed cosmology, with the emphasis on differences between the presence or absence of baryonic physics.

Finally, the UHL PDF \citep{Uhl16,Uhl18} is based on  large deviation theory (LDT), which considers the rate at which probabilities of certain events decay, as a characteristic parameter of the problem varies \citep{Ber16}. Previous key results \citep[eg.,][]{Ber92,Ber94}, concerning the complicated evolution of the large-scale cosmic density field, in which the initial linear fluctuations develop into nonlinear modes due to gravitational instability, have been been connected to LDT. On the other hand, a possible connection between LDT and gravithermodynamics is a novel problem, hence a comparison of the UHL PDF with the GQED predictions is of great interest.

\subsection{The IllustrisTNG Simulations \label{MET}}

The currently publicly available simulations\footnote{For complete information, please refer to the IllustrisTNG project site.} are the TNG300 and TNG100 runs, each with three levels of resolution (with label ``3'' for the lowest resolution runs). These simulations have sizes of: $L_{\rm SIM} ({\rm TNG100}) = 75 ~{\rm Mpc}~h^{-1}$ ; $L_{\rm SIM} ({\rm TNG300}) = 205 ~{\rm Mpc}~h^{-1}$. The TNG300 runs are more adequate for the statistical analysis of large-scale galaxy clustering. The TNG100 follows the same initial conditions (with updated cosmological parameters) of its predecessor, the  Illustris simulation, with volume and resolution lying between the TNG300 and the TNG50 simulations. Hence, the TNG100 is adequate for the study of clustering of at scale of intermediate mass subhaloes. The  TNG300 is more adequate than the TNG100 for the applicability of the GQED, which relies on a gravithermodynamical approach for the description of clustering in the largest structures in the Universe.

We performed a preliminary evaluation to determine the number of CiC computations suitable for our project, given constraints regarding our computational resources and time frame. For the CiC computations, we developed a \cite{Python}  code to implement two extraction methods, given in \cite{Ito88} (hereon, [IIS88]) and \cite{Lei19} (hereon, [Lei19]; details of these methods are presented in the next section). We established a grid of runs considering the lowest resolution versions of the both available volumes, namely,  the TNG100-3 and TNG300-3 (both full  and dark-only runs). In Tab. \ref{TabSim} we present our grid, resulting in a total of $136$ CiC computations.  Note that the comoving number of available subhaloes, understood as gravitationally bound structures identified by the  {\tt SUBFIND} algorithm \citep{Nel19}, decreases for higher redshifts. Details of Tab. \ref{TabSim}  are explained in the next subsections. 

The extraction of the density-in-cells statistics for  the ULH PDF is much more computationally demanding. The methodology involved, to be explained in Sec. \ref{Sec:models}, requires the evaluation of the {\it dark matter density field, directly from the simulation snapshot data on particles} (ie., not from the subhalo catalogue data, as in the case for the CiC statistics).  Given the computational requirements for this procedure, we analyzed the IllustrisTNG 100-3-Dark run at $z=0$ only, using $297$ non-overlapping spheres of radii $10$ Mpc $h^{-1}$, leaving out a margin of $2.5$ Mpc $h^{-1}$ from the simulation box limits. The DM density field in each cell was obtained from the DM snapshot particles, whereas the DM density of subhaloes in each cell was computed from the DM subhaloes catalogue. The IllustrisTNG 100-3 full run at $z =0$ was used to compute the density-in-cells of baryons only, associated with the latter subhaloes.

\begin{table}
\caption{\label{tab-geom} Summary of the analyzed simulations}
%\begin{ruledtabular}
\begin{tabular}{|l|l|l|l|r|r|}
Simulation (TNG) & Redshift & CiC Method & Cell Radii $~[{\rm Mpc}~h^{-1}]$  & $N{\rm sh}$ & $N{\rm excl}$\\ 
(1) & (2) & (3) & (4) & (5) & (6) \\\hline \hline
300-3            & $0.00$   & Rand [IIS88]    & $\mathcal{R}_a = \{10.25, 20.5, 30.75, 41.0 \}$ , $\mathcal{R}_b = \{ 6.0, 12.0, 24.0 \} $  & 391144 & 23535\\
300-3            & $0.05$   & Rand    & $\mathcal{R}_a$ , $\mathcal{R}_b$ & 396402 & 23574\\
300-3            & $0.11$   & Rand    & $\mathcal{R}_a$ , $\mathcal{R}_b$ & 399948 & 23633\\
300-3            & $1.00$   & Rand    & $\mathcal{R}_a$ , $\mathcal{R}_b$ & 429194 & 20056\\
300-3            & $2.00$   & Rand    & $\mathcal{R}_a$ , $\mathcal{R}_b$ & 384814 & 11091\\
300-3            & $3.01$   & Rand    & $\mathcal{R}_a$ , $\mathcal{R}_b$ & 271945 & 4425\\
300-3            & $4.01$   & Rand    & $\mathcal{R}_a$ , $\mathcal{R}_b$ & 156950 & 1233\\
300-3            & $5.00$   & Rand    & $\mathcal{R}_a$ , $\mathcal{R}_b$ & 72591 & 248\\ \hline
300-3-Dark       & $0.00$   & Rand    & $\mathcal{R}_a$ , $\mathcal{R}_b$ & 372177 & 26627\\
300-3-Dark       & $0.05$   & Rand    & $\mathcal{R}_a$ , $\mathcal{R}_b$ & 373988 & 26693\\
300-3-Dark       & $0.11$   & Rand    & $\mathcal{R}_a$ , $\mathcal{R}_b$ & 375860 & 26719\\
300-3-Dark       & $1.00$   & Rand    & $\mathcal{R}_a$ , $\mathcal{R}_b$ & 374473 & 22318\\
300-3-Dark       & $2.00$   & Rand    & $\mathcal{R}_a$ , $\mathcal{R}_b$ & 307526 & 12253\\
300-3-Dark       & $3.01$   & Rand    & $\mathcal{R}_a$ , $\mathcal{R}_b$ & 202685 & 4879\\
300-3-Dark       & $4.01$   & Rand    & $\mathcal{R}_a$ , $\mathcal{R}_b$ & 108238 & 1372\\
300-3-Dark       & $5.00$   & Rand    & $\mathcal{R}_a$ , $\mathcal{R}_b$ & 47491  & 293\\ \hline 
300-3-H          & $0.00$   & Rand ($*$)   & $\mathcal{R}_b$ &  391144 & 23535\\ \hline\hline
100-3            & $0.00$   & Grid [Lei19] & $10.00 , 11.25$ & 118820 & 2221\\
100-3            & $0.00$   & Rand    & $\mathcal{R}^{\prime}_a =  \{ 3.75, 7.5, 11.25, 15.0 \}$  & 118820 & 2221\\
100-3            & $5.00$   & Rand    & $\mathcal{R}^{\prime}_a$ & 68031 & 34 \\ \hline
100-3-Dark       & $0.00$   & Rand    & $\mathcal{R}^{\prime}_a$ &  116020 & 2333 \\
100-3-Dark       & $5.00$   & Rand    & $\mathcal{R}^{\prime}_a$ & 45700  & 36\\ \hline \hline
100-1-Mock       & $0.00$   & Grid     & $10.00$ &4371211 & 4359784 \\
100-1-Mock       & $0.00$   & Rand    & $\mathcal{R}^{\prime}_a$ &  4371211 & 4359784\\ \hline \hline
\label{TabSim}
\end{tabular}
%\end{ruledtabular}
\medskip

{\rm Columns:} (1) simulation label; (2) redshift at which the CiC was applied; (3) applied CiC method; (4) cell radii used for the applied CiC method; (5) total number of candidate subhaloes; and (6) number of excluded subhaloes. The number of cells for the largest selection sphere is $9500$ for the Rand-method, which was doubled (indicated with an asterisk) in the 300-3-H CiC computation. For the Grid-method, a regular grid of $216000$ cells was used.
\end{table}

$~$

%%%%%%%%%%%%%%%%%%%%%%%%%%%%%%%%%%%%%%%%%%%%%%%%%%
\subsection{CiC methods, comoving cell sizes and proxies for galaxies  \label{Sec:proc}}

The CiC statistics are usually extracted from an ensemble of cells with a certain form and size, in which the number of a given class of objects is counted within each cell. In the present case, the objects are the  IllustrisTNG subhaloes and the cells are chosen from different sets of comoving spherical cells. The CiC statistic then reflect the underlying PDF, $f_V(N)$, of the number of objects per cell volume $V$, between $N$ and $N+dN$. The interval or bin $dN$ should be chosen to produce a smooth enough PDF, and this depends on the data sizes. Here we use the terms CiC PDF and $f_V(N)$ interchangeably.

We considered two methods for obtaining the CiC statistics, given by \cite{Ito88}
[IIS88] and \cite{Lei19} [Lei19], which we label as ``Rand-method'' and ``Grid-method'', respectively. The main difference between these methods is the following.

{\em The  Rand-method.} This method is explained in [IIS88], an early numerical investigation about the adequacy of the GQED for describing of the gravitational N-body clustering of $4000$ point particles (each representing a galaxy), evolving in an expanding universe (represented by a sphere of comoving unity radius).  The gravitational interaction of these particles was followed in comoving coordinates, for different cosmological models (total mass density parameter given by $\Omega_m = \{0.01, 0.1, 1.0 \}$, with cold and warm initial velocity distributions). The CiC method in [IIS88] consisted of randomly generating $9500$ points within a selection sphere at the centre of the simulation box. This procedure was repeated for each investigated configuration (i.e., for each snapshot at certain scale factors of interest). The radius of this selection sphere depended on the cell radius for the counting procedure. [IIS88] used selection spheres of radii $R = \{ 0.8, 0.7, 0.6, 0.6\}$ for corresponding cell radii given by $r = \{ 0.1, 0.2, 0.3, 0.4\}$, in which each cell was centred on the randomly generated points. The selection spheres were adopted to avoid boundary effects and to include a sufficient number of particles for the CiC statistics. In other words, a pair $(R_i,r_i)$ defines the selection sphere and cell radius for each computation of $f_V(N)$, where $V \equiv V_i$ refers to the cell volume with radius $r_i$.

{\em The Grid-method.} In  [Lei19], the CiC statistics were obtained for the Illustris TNG100 simulation in the context of 21 cm intensity mapping of neutral hydrogen. The CiC provided the statistics on the mean matter densities of neutral hydrogen, matter, and mass-weighted haloes, in overlapping spheres of comoving radius of $R = 5 $ Mpc $h^{-1}$, centred on a regular $128^3$ grid. This produced $\sim 2$ million density-in-cells samples, in a redshift range of $z = 1$ to $5$. Their fixed choice of $R$ was determined as a compromise between a sufficiently large cell radius and at the same time a large number of independent cells (hence a small enough cell size) for satisfactory statistics (see Sec. 2 in [Lei19] for details).

We performed a preliminary evaluation of the CiC methods using the medium-sized, lowest resolution run, TNG100-3, at $z = 0$. Based on this evaluation, we considered a coarser grid than that in [Lei19] due to our computational constraints, i.e., we used a regular $60^3 = 216000$ grid, imposing an exclusion margin of $750$ kpc to avoid boundary effects. For this preliminary test, we considered two counting cell radii: $10~[{\rm Mpc}~h^{-1}]$ and $11.25~[{\rm Mpc}~h^{-1}]$. For the Rand-method, we used the same number of cells ($9500$) within the selection sphere, applied to the preliminary CiC computations for the TNG100-3 at $z = 0$.  Both CiC methods resulted in qualitatively similar outcomes. Computations using both these methods were also performed for the TNG-100-1-Mock. Given the similarity of the results in both CiC methods, we fixed our full, subsequent analysis to the Rand-method, using the same CiC parameters as in [IIS88].

We summarize two sets of cell radii that we used, as follows:

\begin{enumerate}
\item {{\it The $\mathcal{R}_a$-set:} comoving cell radii as fixed fractions of the simulation size (Rand-method [IIS88]). We defined a set of cell radii given by: $\mathcal{R}_a = \{ 0.1, 0.2, 0.3, 0.4 \} \times R_{\rm SIM} ~[{\rm Mpc}~h^{-1}]$, with $R_{\rm SIM} =  L_{\rm SIM}/2$. Each cell centre in set $\mathcal{R}_a$ must be inside a selection radius given, respectively, by: $R_{\rm{sph}} = \{ 0.8, 0.7, 0.6, 0.6 \} \times R_{\rm SIM} ~[{\rm Mpc}~h^{-1}]$, to avoid boundary effects. This gives different comoving cell sizes for the TNG300-3 and TNG100-3 runs, namely: for TNG300-3, $\mathcal{R}_a = \{10.25, 20.5, 30.75, 41.0 \}  ~[{\rm Mpc}~h^{-1}]$; and for TNG100-3,  $\mathcal{R}^{\prime}_a = \{ 3.75, 7.5, 11.25, 15.0 \}  ~[{\rm Mpc}~h^{-1}]$. }
\item {{\it The $\mathcal{R}_b$-set:} comoving cell radii of $\mathcal{R}_b = \{ 6.0, 12.0, 24.0 \} ~[{\rm Mpc}~h^{-1}]$. This set was only analyzed for the TNG300-3 simulations. This choice of radii is the same as that explored in \cite{Yan11}. The respective selection radii were defined as: $R_{\rm{sph}} = \{ 0.8, 0.7, 0.6\} \times R_{\rm SIM} ~[{\rm Mpc}~h^{-1}]$.}
\end{enumerate}

We used the {\tt SUBFIND} object catalogue \citep[cf. ][and references therein]{Pil18} for the identification of  subhaloes the counting procedure. We selected subhaloes (as proxies for galaxies) with attributes restricted to the following criteria:

\begin{itemize}
\item{Subhalo mass range: subhaloes had a minimum mass of $2.5 \times 10^8 ~{\rm M}_{\odot}~h^{-1}$ (as in [Lei19]), and a maximum cutoff mass of $\sim 10^{13} ~{\rm M}_{\odot}~h^{-1}$.}
\item{Subhalo attribute `SubhaloFlag' had to be equal to $1$  (indicating that the particles composing the subhalo have a cosmic origin).}
\end{itemize}

We do not differentiate between central and satellite galaxies \citep{Pil18}. For the dark-only simulations, we used the same criteria as above. The overall matching between bound structures in the {\tt SUBFIND} catalogue in both full and dark-only runs is the basis for a direct comparison between their corresponding CiC statistics results. In Tab. \ref{TabSim}, we quote in columns (5) and (6) the total number of initial candidate subhaloes ($N{\rm sh}$) and the number of excluded subhaloes ($N{\rm excl}$) after applying the criteria above. 

As mentioned previously, we used different criteria for galaxy proxies for the TNG-100-1 simulation,  which was based on the publicly available mock catalogue by \cite{Rod19} (TNG-100-1-Mock). In this case, for selecting admissible subhaloes, we matched the individual subhalo ID's listed in that catalogue to those in the  {\tt SUBFIND} catalogue of the TNG100-1 at $z=0$. The mock catalogue includes subhaloes with (total) stellar mass greater that  $10^{9.5} M_{\odot}$ and tailored to the Sloan Digital Sky Survey  \cite[SDSS][]{Yor00}, as explained in \cite{Rod19}.  Furthermore, we cut magnitudes in the $r$-band for synthetic objects fainter than $17.77$ mag, resulting in $11427$ subhaloes, representing galaxies as would be observed in that band at $z \lesssim 0.05$.

%%%%%%%%%%%%%%%%%%%%%%%%%%%%%%%%%%%%%%%%%%%%%%%%%%
\subsection{Models and fitting procedure \label{Sec:models}}

\subsubsection{The GQED (CiC PDF):}

The gravithermodynamical theory leading to the simplest form of the GQED assumes that the galaxy clustering evolves through a series of quasi-equilibrium states due to a cancelling effect of the long range, mean gravitational field, from the expansion of the Universe \citep{SaslawBOOK00}. The resulting CiC PDF, $f_V(N) \equiv f_{\rm GQED}(N) $, representing the probability that a cell of volume $V$ of arbitrary shape contains $N$ galaxies, has a form which modifies the Poisson PDF due to the presence of correlations:

\begin{equation}
f_{\rm GQED}(N) = 
{ \bar{N} (1-b) \over N!} 
\left [ \bar{N} (1-b) + Nb \right ] ^{N-1}
\exp \left [ -\bar{N}(1-b)-Nb\right ], \label{GQED}
\end{equation}

\noindent where $\bar{N} = nV$ is the expected average number of galaxies in the given volume $V$, with  average density $n$.  The aggregation parameter $b$ is related to the degree of clustering of galaxies at a certain state, and represents the average departure from a noninteracting ensemble of cells of a given volume at that state. A more detailed physical interpretation for this parameter, in terms of the average density and kinetic temperature $T$ of the system, was explored in subsequent developments \citep[e.g.,][]{Sas96,Ahm02}. Numerical studies of the gravitational clustering of galaxies in a range of expanding universe models have qualitatively shown that structures quickly relax to the GQED form of Eq. \ref{GQED}, and that they subsequently evolve, in general, through a series of quasi-equilibrium states \citep[e.g.][]{Ito88, Ito93, WKS20}. Each of such states would satisfy Eq. \ref{GQED} for a given value of $b$, in other words, the theory admits a time-dependent $b(t)$, which increases slowly from a lower value, as clustering proceeds hierarchically into larger and larger scales with time \citep{Sas86,SaslawBOOK00}.

It is also important to understand how the value of $b$ depends on the counting cell size \citep[eg.,][]{She96}, and how correlations evolve more rapidly, depending on the spatial scale, as nonlinear structures develop first from near-neighbour interactions on smaller scales than in larger ones. This effect is explicitly encoded in the expression for $b$ in terms of the two-point correlation function $\xi$ \citep[e.g.,][]{Sas84,SaslawBOOK00}:

\begin{equation}
b = -{W \over 2K} = {2 \pi G m^2 n \over 3 T} \int_0^{\infty} \xi(n,T,r) r dr, \label{B}
\end{equation}
\noindent where $m$ is the average galaxy mass. The virial ratio above actually represents an ensemble average ratio of the gravitational {\it correlation} energy ($W$) to twice the kinetic energy ($K$) of the {\it peculiar velocities} of galaxies in an idealized {\it infinite} universe. Depending on the initial conditions, there could be a characteristic scale above which the correlation function would not contribute to the integral in Eq. \ref{B}, leading to $b \rightarrow 0$ (Poisson). For very small scales, with either one or zero galaxies, the distribution function would also tend to a Poissonian one \citep{SaslawBOOK00}.

\subsubsection{The NBD (CiC PDF): }

The NBD was initially introduced in cosmology without an underlying physical foundation by \cite{Car83}, as a good approximation to the distribution of Zwicky clusters, and further developed by \cite{Eli92}. The NBD can be expressed as:

\begin{equation}
f_{\rm NBD}(N) = 
{\Gamma \left ( N + {1 \over g} \right )
\over
\Gamma \left ( {1 \over g} \right ) N!}
{\bar{N}^N \left ( {1 \over g} \right ) ^{1 \over g}
\over
\left ( \bar{N} + {1 \over g} \right ) ^{N + {1 \over g}}}, \label{NBD}
\end{equation}

\noindent where $g$ is the NBD clustering parameter, which is also equivalent to the two-point correlation function, $\xi(V)$ \citep[eg.,][]{WKS20}.

\subsubsection{The fitting procedure to the CiC PDFs}

We directly fit the GQED and NBD models with the respective   $(\bar{N}, b, g)$ as free parameters.  The fittings were performed using the  {\tt curve\_fit} module in the {\tt scipy} library \cite{Scipy}. We chose the {\tt lm} optimization method (Levenberg-Marquardt algorithm) in the case of the GQED model fitting, and the  {\tt trf} (Trust Region Reflective algorithm, with parameter bounds set to $0$ and $\infty$) for the case of the NBD model fitting, which has proven necessary for achieving convergence. For handling large values of N for the factorial evaluation, we used the Python module {\tt Decimal}. The fittings were performed on CiC histograms with variable bins widths, $dN$, which mainly depended on the cell size. After a series of tests, the bin widths also had to be adjusted for different redshifts, because of the size and spread of the CiC population as a function of redshift. The bin widths $dN$, as a function of redshift and cell radius, were:

\begin{itemize}
\item{For TNG 100-3 and TNG 100-1-Mock:}
\subitem{$dN_{\mathcal{R}^{\prime}_a}(z=0) = \{ 1, 10, 25, 50\}$, and}
\subitem{$dN_{\mathcal{R}^{\prime}_a}(z=5) = \{ 1, 5, 10, 10\}$.}
\item{For TNG 300-3:}
\subitem{$dN_{\mathcal{R}_a}(z \leq 3.01) = \{ 1, 10, 30, 50\}$,}
\subitem{$dN_{\mathcal{R}_a}(z > 3.01) = \{ 1, 1, 1, 2\}$; and}
\subitem{$dN_{\mathcal{R}_b}(z \leq 3.01) = \{ 1, 3, 10\}$, }
\subitem{$dN_{\mathcal{R}_b}(z > 3.01) = \{1, 1, 1\}$. }
\end{itemize}

\noindent In the notation above, bin widths (inside brackets) are listed in the same order as the corresponding sets of cell radii (cf. Sec. \ref{Sec:proc}).

$~$

\subsubsection{The dGQED (density-in-cells PDF): }

We derived the matter density-in-cells dGQED PDF, $\mathcal{P}_{\rm GQED}(\rho)$, corresponding to the GQED, Eq. \ref{GQED}, under some simplifying assumptions \citep[our Ansatz follows from a similar procedure derived for the velocity PDF, $f(v)$, as explained in][]{SaslawBOOK00}. First, we consider that fluctuations in the counting number $N$ (per fixed cell of volume $V$), among cells in the ensemble, correspond to fluctuations in mass density, so that:

\begin{equation}
\rho = \alpha {N \langle m \rangle \over V},
\end{equation}

\noindent where $\langle m \rangle$ is the average expected value for individual galaxy masses under a uniform Poisson distribution, and $\alpha$ is a factor representing local departures from a uniform distribution. We make the simplifying assumption that $\alpha$ is a constant, given by its average value over the entire ensemble. We then rescale Eq. \ref{GQED} from number fluctuations to matter density fluctuations by replacing $N$ with $N \langle m \rangle = \rho V/\alpha$, and the average counting number in the ensemble, $\bar{N}$, with $\bar{N} \langle m \rangle = \bar{\rho} V/\alpha$, where $\bar{\rho}$ is the average matter density-in-cells in the ensemble. Next, we use the identity $N! = \Gamma(N+1)$, making the continuous replacement for $N$. Finally, we convert $f_{\rm GQED}(N) \Delta N \mapsto \mathcal{P}_{\rm GQED}(\rho) d\rho$ using: $\Delta N \mapsto \beta d\rho$, with $\beta \equiv V/\alpha$. The resulting PDF is given by:

\begin{equation}
\mathcal{P}_{\rm GQED}(\rho) = 
{ \beta^2 \bar{\rho} (1-b) \over \Gamma(\beta\rho+1)} 
\left \{ \beta \left [  \bar{\rho} (1-b) + \rho b \right ] \right \} ^{\beta \rho-1}
\exp \left \{-\beta  \left [ \bar{\rho}(1-b) +\rho b\right ]  \right \} 
~~~~~~~~{\rm [dGQED ~PDF]}.\label{GQEDPDF}
\end{equation}

\begin{figure}
\centering
\includegraphics
[trim= 0in 3.5in 0in 3.5in,clip,width=1.0\linewidth]{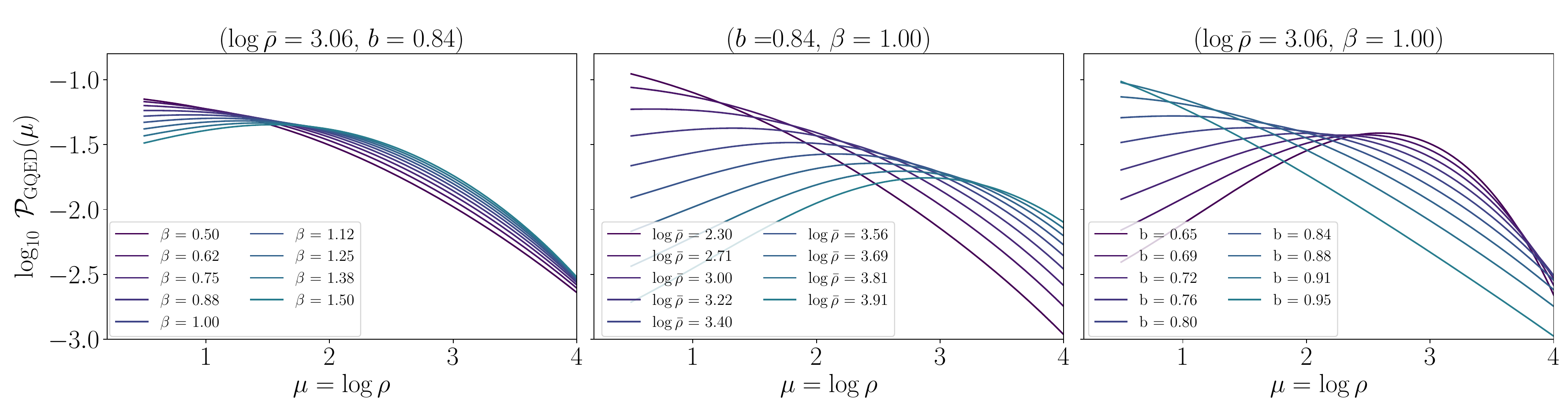} 
\caption{  (Colour online). Behaviour of Eq. \ref{GQEDPDF} under variations in $\beta$, $\bar{\rho}$ and $b$ (from left to right panels, respectively). The titles over each panel indicate the values of the quantities which were held fixed. \label{DiC_PDF_GQED}}
\end{figure}

The  dGQED PDF is presented in Fig. \ref{DiC_PDF_GQED}, in which we illustrate its behaviour under variations in $\beta$, $\bar{\rho}$ and $b$. In particular, the behaviour of the dGQED in the latter panel (variations in $b$) is qualitatively similar to the GQED CiC PDF, as can be seen in the panels shown in Fig. 28.1 of \cite{SaslawBOOK00}.

We also derive an approximate relation between the density-in-cells variance and the $b$ parameter for the dGQED. From Eq. (26.34) and subsequent discussions in the Chap. 28 of \cite{SaslawBOOK00}, the variance of fluctuations for the number of objects in a cell, in the case of the GQED, is given by: $\sigma^2 \approx \bar{N}(1-b)^{-2}$ (to order $\bar{N}^{-1/2}$). In our Ansatz, we propose:
$\sigma_{\rho}^2 \approx \bar{\rho}(1-b)^{-2}$. Then, using the approximate correspondence of variances in log-densities and densities, ${\rm Var}[\log \rho] \approx {1 \over \bar{\rho}^2} {\rm Var}[\rho]$ (where the expectation value is $E[\rho] = \bar{\rho}$), we find:

\begin{equation}
\sigma_{\log{\rho}}^2 \equiv \sigma_{\mu; \rm dGQED}^2 \approx 
{1 \over \bar{\rho}}{1 \over (1-b)^2}. \label{sigmadGQED}
\end{equation}

$~$

\begin{figure}
\centering
\includegraphics[width=1.0\columnwidth]{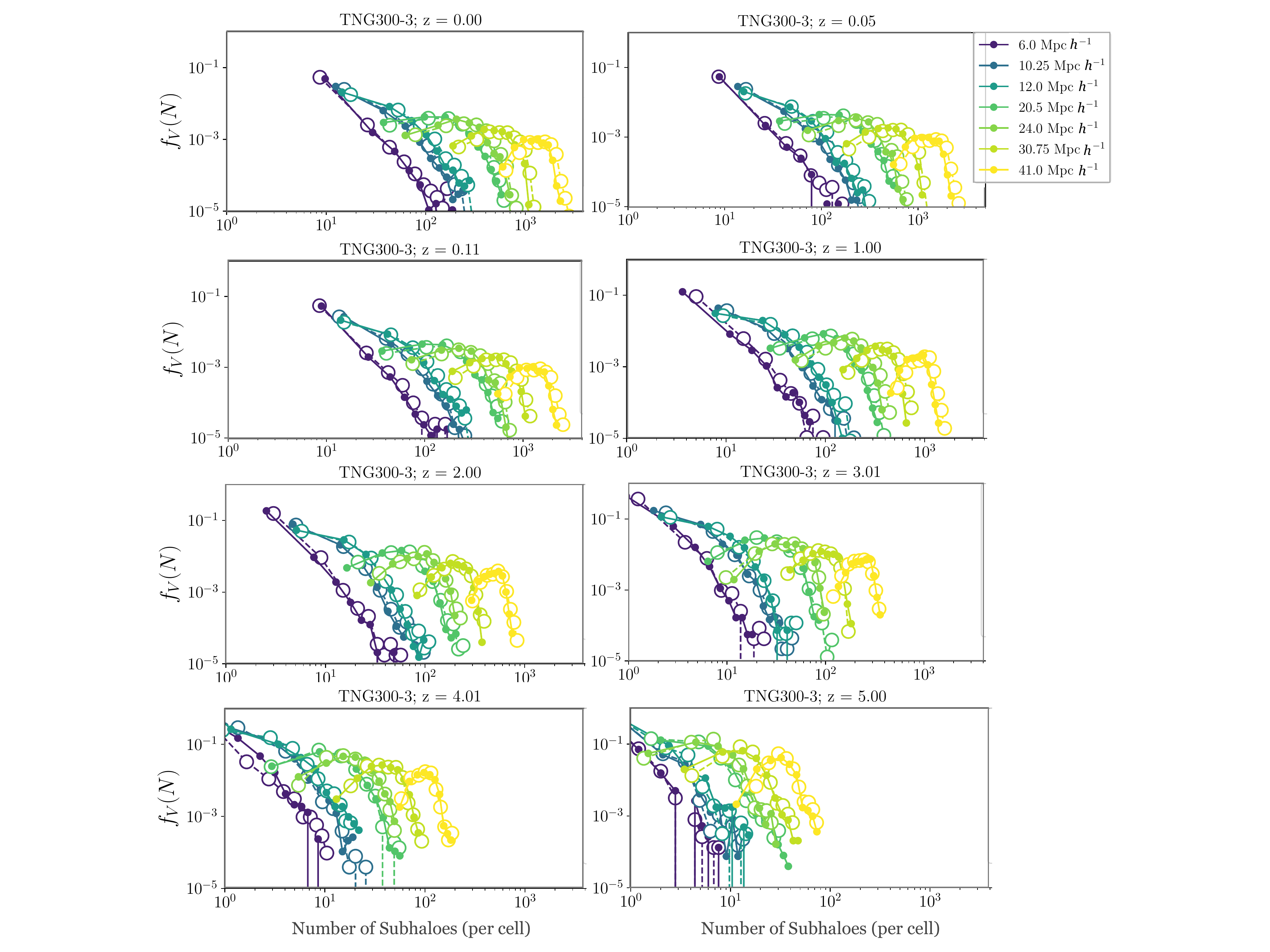} 
\caption{ (Colour online).  Normalized CiC histograms (in log-log scale) for the TNG-300-3 simulations, starting from redshift $z= 0.0$ (top, left) to $z= 5.00$ (bottom, right), for comoving cell sizes indicated in the legend (covering both $\mathcal{R}_a$ and $\mathcal{R}_b$ sets, cf. Tab. \ref{TabSim}). Darker colours are used for smaller radii. The number of bins in this representation is fixed to $N_{\rm bins} = 10$. Symbols are used instead of histogram bars: smaller (filled) circles connected with continuous lines for the full  runs; larger (empty) circles connected with dashed lines for the corresponding dark-only runs.   \label{FVN-TNG300-3}}
\end{figure}

\subsubsection {The UHL (density-in-cells PDF):}

The UHL PDF is based on a bias model, relating the density-in-cells statistics of the dark matter field, $\rho_{\rm m}$, and those of the dark matter subhaloes, $\rho_{\rm sh}$ \citep{Uhl16, Uhl18}.  Here we use the saddle-point approximation for the integral form of the PDF, as explained in \cite{Uhl16}. Up to the mildly nonlinear clustering regime ($\sigma_{\mu}^2 \lesssim 1$), the UHL PDF for $\rho_{\rm m}$ within a sphere of radius $R$ and redshift $z$ is given by:

\begin {equation}
\mathcal{P}_R(\rho_{\rm m}) = 
\sqrt{
{\Psi_R^{\prime \prime}(\rho_{\rm m}) + \Psi_R^{\prime}(\rho_{\rm m})/\rho_{\rm m}
\over
2 \pi \sigma^2_{\mu}
}
}
\exp 
\left (
-{ \Psi_R(\rho_{\rm m})
\over
\sigma_{\mu}^2}
\right ), \label{DMPDF}
\end {equation}

\noindent with:

\begin{equation}
\Psi_R (\rho_{\rm m}) =
{ \tau_{\rm SC}^2(\rho_{\rm m}) \sigma_{\rm L}^2(R)
\over
2 \sigma_L^2(R\rho_{\rm m}^{1/3})
},
\end{equation}

\begin{equation}
\tau_{\rm SC}(\rho_{\rm m}) = \nu(1 - \rho_{\rm m}^{-1/\nu}),
\end{equation}

\noindent where we used $\nu = 21/13$; the prime in Eq. \ref{DMPDF} denotes a derivative with respect to $\rho_{\rm m}$,  $\sigma_{\mu}$ is the nonlinear variance of the corresponding log-density ($\mu = \log \rho$); $\tau_{\rm SC}$ is the linear density contrast averaged within the Lagrangian radius \citep[see details in ][]{Uhl18}; $\sigma_L$ is the linear variance of the density field on a scale $R$. For the latter, we used {\tt sigma} module of the Colossus Python Toolkit for cosmological calculations \citep{COL}, giving $\sigma_L^2 = 0.486$. We fit the data to the quadratic bias model in the log-densities, given by:

\begin{equation}
\mu_m = b_0+b_1 \mu_{\rm sh} + b_2 \mu_{\rm sh}^2. \label{bias}
\end{equation}

\noindent The best-fit parameters provided the mean relation $\mu_m(\mu_{\rm sh})$ of the bias model, and the subhalo PDF $\mathcal{P}_{\rm sh}$ was obtained from the dark matter PDF $\mathcal{P}_m$ (Eq. \ref{DMPDF}) by conservation of probability:

\begin{equation}
\mathcal{P}_{\rm Uhl} \equiv 
\mathcal{P}_{\rm sh} (\rho_{\rm sh})= \mathcal{P}_m [ 
\rho_{\rm m} (\rho_{\rm sh})
]
\left | {d\rho_{\rm m}\over d\rho_{\rm sh}} 
\right | ~~~~~~~~{\rm [UHL ~PDF].}\label{PDFSH}
\end{equation}

We also used the large density tail ($\rho \gg 1$) of the UHL PDF, as given  in \cite{Uhl16} (their Eq. (27)):

\begin{equation}
\mathcal{P}_{\rm tail}(\rho_{\rm m}) \rightarrow
{
(n_{\sigma}+3)\nu
\over
6\sqrt{\pi\sigma^2_{\mu}}
}
\exp
\left [
-{
\nu^2(\rho_{\rm m}^{1\over \nu} -1)^2
 \rho_{\rm m}^{{n_{\sigma}+3 \over 3} - {2 \over \nu}}
\over
2 \sigma^2_{\mu}
}
\right 
]
\rho_{\rm m}^{{n_{\sigma}-3 \over 6}} 
, \label{TAIL}
\end{equation}

\noindent where we used the notation $n_{\sigma}$ for the index of a power-law initial power spectrum, occurring in their Eq. (13); namely, the variance of the field fluctuation within a sphere of radius $R$ follows the relation:

\begin{equation}
\sigma^2(R) = \sigma^2(R_p)\left ( R/R_p \right )^{-(n_{\sigma} +3)},
\end{equation}
\noindent where $R_p$ is a pivot scale (see their paper for details).

We developed a Python code to implement these calculations. The positivity condition involving the derivatives of the $\Psi$ function in Eq. \ref{DMPDF} was met in the ranges analyzed; hence the saddle-point approximation was adequate. We checked the validity of our code by testing against data ranges and parameters found in \cite{Uhl18}.  For a set of UHL PDFs computed with a different code ({\tt LSSFAST}), for different variances and radii, see \cite{Cod16}.

$~$

%%%%%%%%%%%%%%%%%%%%%%%%%%%%%%%%%%%%%%%%%%%%

% trim 0 inches on the left, 1.3 inches on the bottom, 0 inches on the right, and 1.3 inches on the top. 

\begin{figure}
\centering
\includegraphics[trim= 0in 2.5in 0in 1.5in,clip,width=1.0\linewidth]{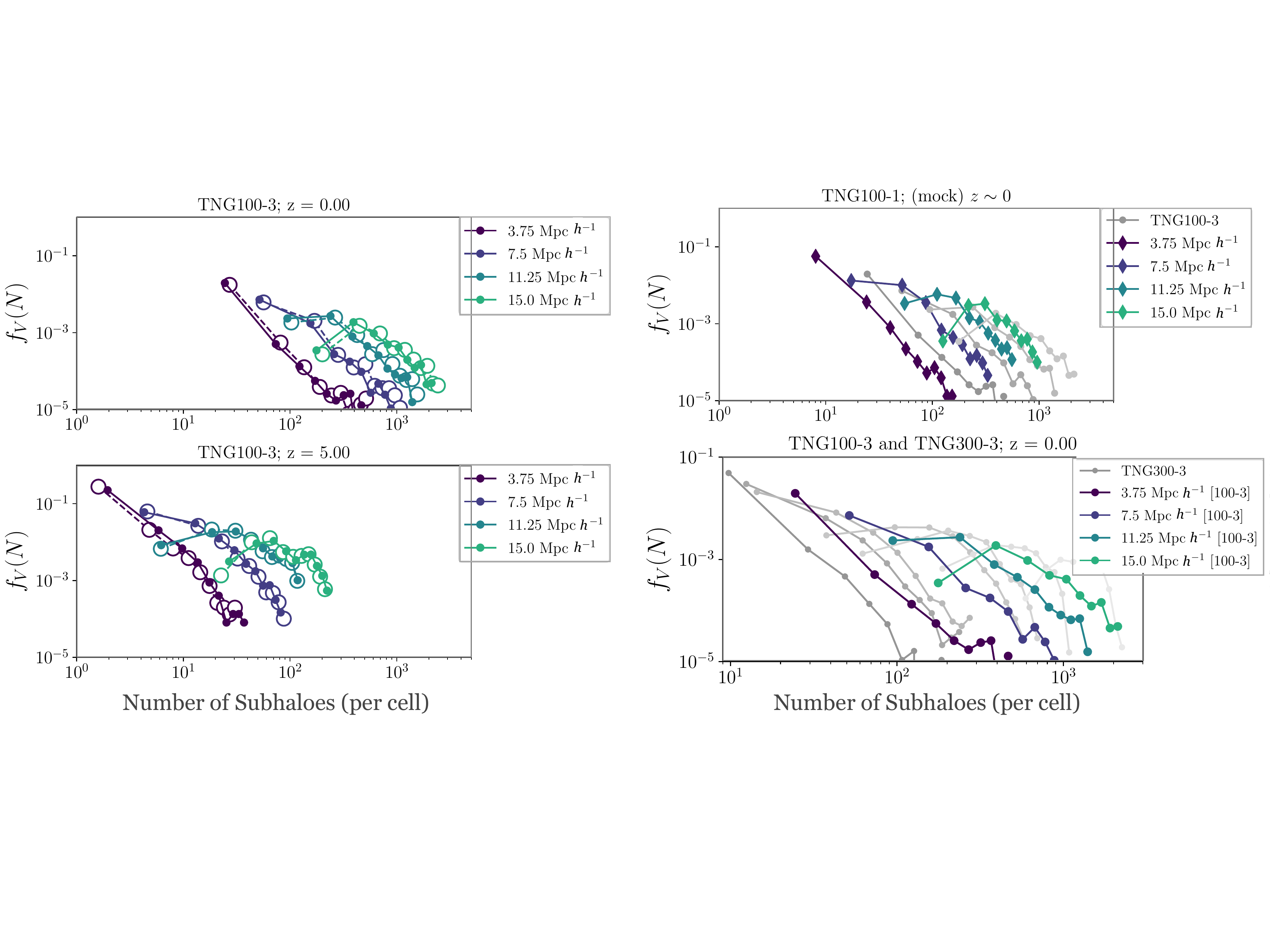} 
\caption{ (Colour online).  Normalized CiC histograms (in log-log scale) for the TNG-100-3 simulations. {\it Left panels:}  redshift $z= 0.0$ (top)  and $z= 5.00$ (bottom). {\it Right panels:} in the top panel, the TNG-100-1-Mock run, with diamond symbols, whereas filled circles in grayscale represent the results for TNG-100-3 full  at $z = 0.0$; in the bottom panel, the TNG100-3 full distributions are compared with those of the TNG300-3 full runs (both at $z = 0.0$), showing the effects of volume size. \label{FVN-TNG100-3}}
\end{figure}

\begin{figure}
\centering
\includegraphics[trim= 0in 1.3in 0in 0.5in,clip,width=1.0\linewidth]{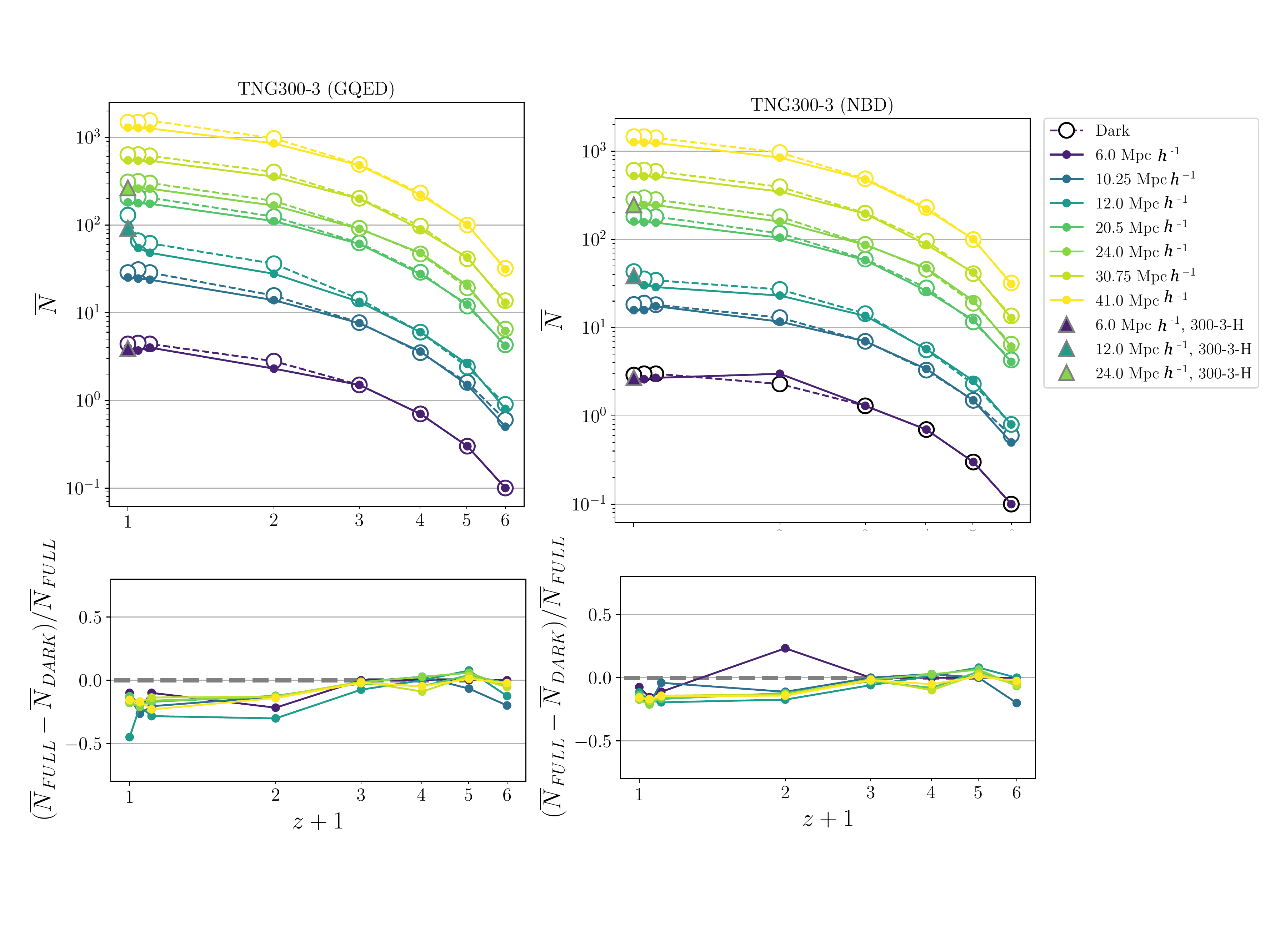} 
\caption{ (Colour online). Results for the TNG300-3 simulations, redshifts $z = \{0.00, 0.05, 0.11, 1.00, 2.00, 3.01, 4.01, 5.00 \}$, for various comoving cell radius (darker colours for smaller radii, see legend). Filled circles connected with continuous lines refer to the full simulation runs; larger, empty circles connected with dashed lines refer to the respective dark-only simulation runs. The TNG300-3-H (full) runs are represented by triangles at $z = 0$.
{\em Top panels:} behaviour of the average number of subhaloes per cell, $\bar{N}$, as a function of $z$, obtained from the fit to the GQED (left panel) and the NBD (right panel) models. {\em Bottom panels:} fractional  difference between full and dark-only results, relatively to the full  runs (GQED: left panel; NBD: right panel). \label{PARAMS300-N}}
\end{figure}

\begin{figure}
\centering
\includegraphics[trim= 0in 1.5in 0in 0.5in,clip,width=1.0\linewidth]{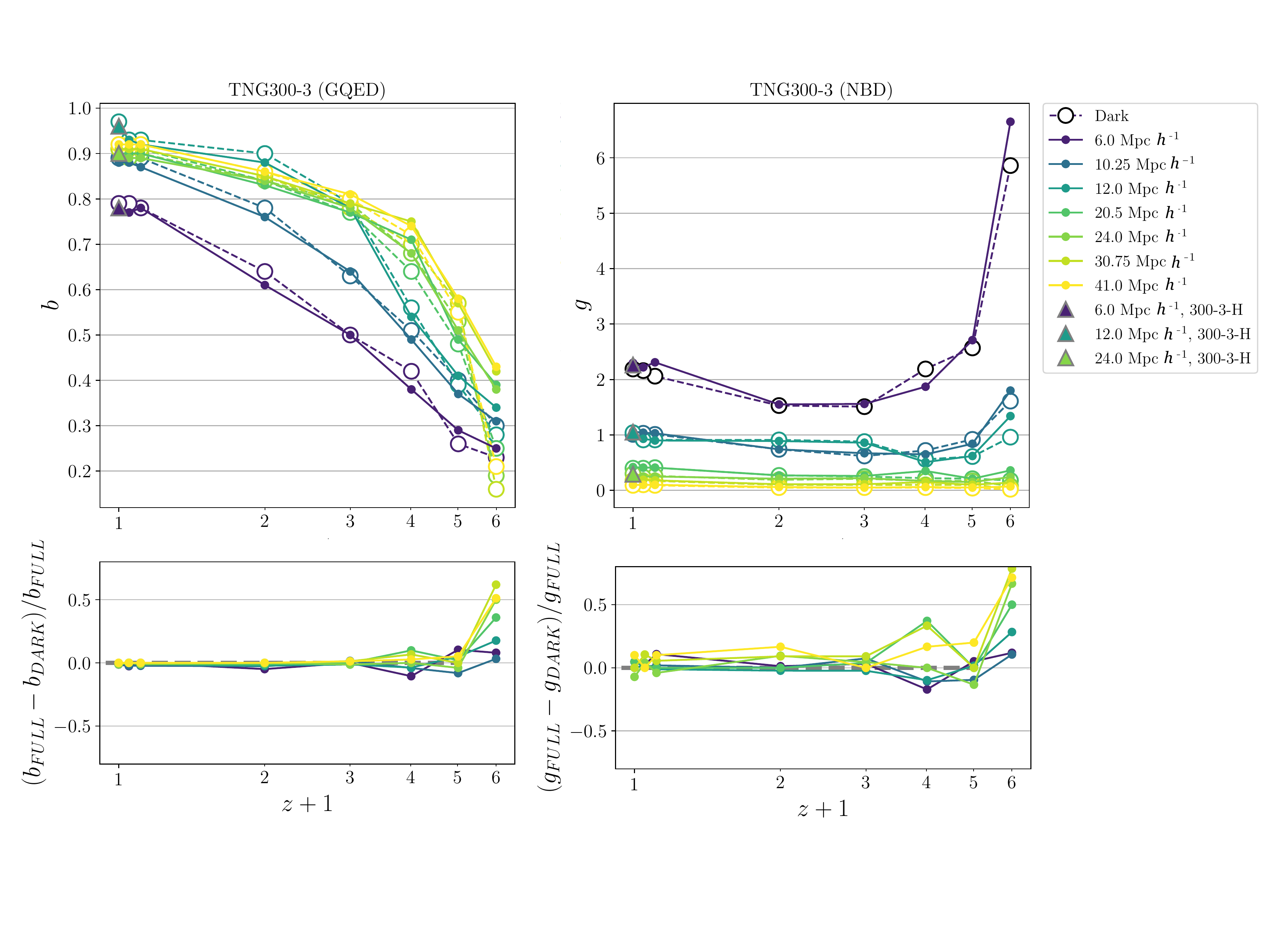}
\caption{ (Colour online). 
Results for the TNG300-3 simulations, redshifts $z = \{0.00, 0.05, 0.11, 1.00, 2.00, 3.01, 4.01, 5.00 \}$. {\em Top panels:} behaviour of the clustering parameter as a function of $z$, obtained from the CiC PDF fit to the GQED ($b$ parameter; left panel) and NBD ($g$ parameter, right panel) models, for various comoving cell radius, as indicated in the legend. {\rm Bottom panels:} fractional  difference between full and dark-only results, relatively to the full  runs (GQED: left panel; NBD: right panel). Symbols are the same as in the previous figure.\label{PARAMS-TNG300-3-clust}}
\end{figure}

%%%%%%%%%%%%%%%%%%%%%%%%%%%%%%%%%%%%%%%%%%%%%%%%%%
% 3 - RESULTS
%%%%%%%%%%%%%%%%%%%%%%%%%%%%%%%%%%%%%%%%%%%%%%%%%%
\section{Results \label{RES}}

%%%%%%%%%%%%%%%%%%%%%%%%%%%%%%%%%%%%%%%%%%%%%%%%%%
\subsection{Qualitative presentation of the CiC distributions \label{Sec:FVN}}

Following a similar display as in \cite{WKS20}, we present all our normalized CiC PDFs (in log-log scale), in Figs \ref{FVN-TNG300-3} and \ref{FVN-TNG100-3} for the TNG300-3 and TNG100-3 runs, respectively. We also present the TNG-100-1-Mock results (Fig. \ref{FVN-TNG100-3}, top right panel)  with TNG100-3 full results together in the same panel for direct comparison. Finally, we present in Fig. \ref{FVN-TNG100-3}, bottom right panel, a comparison between TNG300-3 and TNG100-3, for the  full runs at $z = 0$. The resulting CiC distributions agree with previous numerical results in the literature, even though being purely gravitational ones \citep[e.g.][]{Ito88, Ito93,Yan11, HGil17,WKS20}.

Full and dark-only simulations follow remarkably similar CiC distributions, in both the TNG300-3 and TNG100-3 runs. There are, however, differences which can be finely traced by a quantitative analysis of the best-fit parameters, to be discussed in Sec. \ref{Sec:params}. Both simulations volumes cover (approximately) similar $N$ ranges in their CiC PDFs, even though the comoving cell sizes are different. All CiC distributions tend to gradually spread towards $z = 0$  as more subhaloes are formed and the gravitational clustering evolves. The height of the CiC PDF peak decreases for larger cell sizes due to spreading and normalization, even considering a larger number of samplings. The height of the CiC PDF peak decreases for lower redshifts, as the distribution spreads into a wider range of $N$. Note, however, that in \cite{WKS20} (cf. their Fig. 2), the trend goes in opposite direction: the peak height decreases at higher redshifts. as the distribution spreads into a wider range of $N$ for {\em physical} cell sizes, as explained in their paper. The reason for this difference appears to be due to a different measurement criterium for the counting cells. In \cite{WKS20}, the authors use physical cell sizes, whereas we use comoving cell sizes. In the former case, the cell encloses a larger volume at higher redshifts, whereas we follow the same comoving volume.

Our statistics are necessarily poorer for higher redshifts and smaller cell sizes, as the total number of subhaloes available for the comoving counting (gravitationally bound structures) is less than that at $z=0$ (cf. Tab. \ref{TabSim}, column $N_{\rm sh}$). An analysis of the cell sampling showed that, considering the TNG300-3-H case, doubling of the number of counting cells leads only to a linear increase on the resulting $N_{\rm samples}$ (at least for $z=0$). We conclude that the CiC statistics for the smallest cell size would require at least an order of magnitude increase in the number of counting cells, so our results for the smallest cell size in both TNG100-3 and TNG300-3 runs should be considered less certain. 
In  \cite{WKS20}, the CiC distribution was found to be less smooth in larger cells at higher redshifts; we attribute this opposite effect given the physical (theirs) vs. comoving (ours) counting cell method.

%%%%%%%%%%%%%%%%%%%%%%%%%%%%%%%%%%%%%%%%%%%%%%%%%%
\subsection{Quantitative analysis of the best-fit parameters \label{Sec:params}}

\subsubsection{CiC best-fit parameters}

In Fig. \ref{PARAMS300-N}, we show the best-fit average number of subhaloes per cell as a function of redshift, for the TNG300-3 runs (GQED: left panel; NBD: right panel).  The full and dark-only runs follow closely the same overall behaviour of this parameter. 
In order to see more clearly distinctions between the full and dark-only results, we show in the bottom panels of Fig. \ref{PARAMS300-N} the fractional difference between these runs, relatively to the full runs, as a function of redshift, for the given cell sizes. At lower redshifts, the  $\bar{N}$ parameter tends to be systematically higher for the dark-only runs. This effect is small ($\lesssim 5 \%$) but clearer in the GQED case. 

We present in Fig. \ref{PARAMS-TNG300-3-clust} the results for the TNG300-3 runs in terms of the clustering parameters $b$ (GQED; left panel) and $g$ (NBD; right panel), as functions of redshift and cell size, in the same format as in Fig. \ref{PARAMS300-N}. For the GQED, we see a clear trend, for both  full and dark-only  runs and for all cell sizes, showing smaller values of $b$ at higher redshifts to greater values of $b$ at lower redshifts.  Smaller cells (at fixed redshift) tend to show smaller values of $b$. For the NBD, $g$ values are generally larger for smaller cell sizes. The bottom panels of Fig. \ref{PARAMS-TNG300-3-clust} show the fractional difference between full and dark-only $b$  (GQED) and $g$ (NBD) results, relatively to the full runs, as a function of redshift, for the given cell sizes. Overall, both full and dark-only runs present a remarkably similar evolution of their respective clustering parameters at redshifts $z \lesssim 2$, with greater values of $b$ and $g$, for the full runs and for larger cell sizes at $z=5$.

For the TNG100-3, we have only analyzed the  $z = \{0.00, 5.00 \}$ runs, and  for brevity we omit related figures. The best-fit parameters follow in this case the same overall trends in terms of comoving cell size as in the TNG300-3 runs, at the $z = 0$ and $z=5$ values. The main differences between these simulation volumes are the following. For the $\bar{N}$ parameter, in both GQED and NBD, the TNG100-3 values at $z=5$ are higher than those of the TNG300-3 runs, but the increase of this parameter towards the values at $z=0$ are less than one order of magnitude, that is, a lower relative increase than in the TNG300-3 runs. The $b$ parameter shows higher values at $z=5$ as compared to the TNG300-3 runs, but converges to those of the TNG300-3 runs at $z=0$.

%%%%%%%%%%%%%%%%%%%%%%%%%%%%%%%%%%%%%%%%%%%%%%%%%%
\subsubsection{CiC: IllustrisTNG and observations \label{Sec:obs}}

\begin{figure}
\centering
\includegraphics[width=1.0\columnwidth]{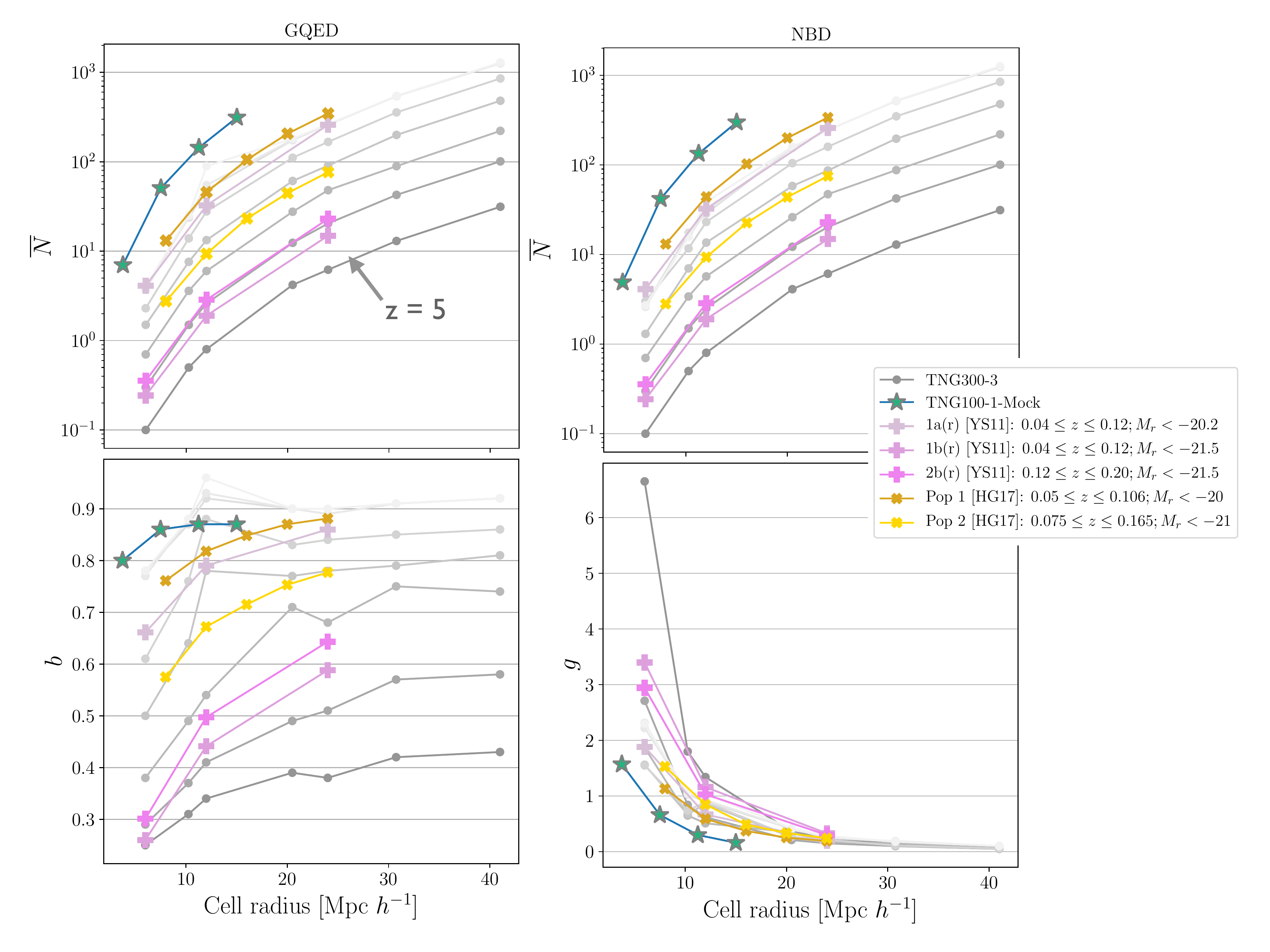} 
\caption{ (Colour online). Behaviour of the best-fit parameters for the TNG300-3 runs as a function of comoving cell size; full circles connected with thin lines are shown with greyscale tones such that the darker tones represent greater redshifts ($z=5$ results are indicated for reference). 
Also shown are the TNG100-1-Mock results, as well as those for the corresponding best-fit values for the data in [YS11], in terms of the 1a(r), 1b(r) and 2b(r) samples, are indicated. Also shown are results for the data in [HG17], for the population 1 and 2 samples.  Redshift and $M_r$ magnitudes ranges for those data are indicated in the legend.
{\em Top panels}: behaviour of the average number of subhaloes per cell, $\bar{N}$, as a function of comoving cell size, obtained from the CiC PDF fit to the GQED (left) and NBD (right) models.  {\em Bottom panels:} behaviour of the best-fit clustering parameters, as a function of comoving cell size: $b$ parameter (GQED, left); and $g$ parameter (NBD, right).  
\label{PARAMS-TNG300-3-BarNbg-RADII}}
\end{figure}

We compare our results with those obtained by  \cite{Yan11} (hereon [YS11]) and  \cite{HGil17} (hereon [HG17]). In [YS11], the galaxy catalogue is a flux-limited ($r < 17.6$) subsample taken from Sloan Digital Sky Survey (SDSS)  \cite[SDSS][DR7]{SDSS7}, with additional absolute magnitude cuts, resulting in $3$ subsamples within two redshift ranges, namely:  
{\em 1a(r):} $0.04 \le z \le  0.12$, $M_r < -20.2$; 
{\em 1b(r):} $0.04 \le z \le  0.12$, $M_r < -21.5$; 
{\em 2b(r):} $0.12 \le z \le  0.20$, $M_r < -21.5$. The lower cut at $z \le 0.04$ excludes the Coma and Virgo clusters; the subsample follows the Hubble flow. On the other hand, the higher redshift range includes the SDSS great wall, allowing a comparison of potential differences between both ranges. In [HG17], the data were also based on the DR7, with galaxy catalogue provided by The New York University - Value Added Galaxy Catalog \citep{Bla05}. They also used the LasDamas simulation catalogue \citep{LasDamas} to estimate the uncertainties in the resulting CiC distribution. Their selected samples are given by two populations: 
{\em Pop 1:} $0.050 \le z \le  0.106$, $M_r < -20.0$; 
{\em Pop 2:} $0.075 \le z \le  0.165$, $M_r < -21.0$; which are placed roughly within samples 1a(r) and 2b(r) from [YS11], respectively.

In Fig. \ref{PARAMS-TNG300-3-BarNbg-RADII}, we present the behaviour of the best-fit parameters $\bar{N}$, $b$ and $g$  for the TNG300-3 runs, as a function of comoving cell size, for all redshifts analyzed (darker greyscale tones indicate greater redshifts, with the $z=5$ results indicated for reference). Also shown are the TNG100-1-Mock results, and the results from [YS11] and [HG17]. All samples (observational and simulated) follow similar behaviours in terms of cell size, for both GQED and NBD models, only differing in terms of relative amplitudes in  their best-fit parameters. Trends  in $b$ and $g$ are more heterogeneous across the samples and simulation runs than those found in $\bar{N}$.

\subsubsection{Density-in-Cells}

\begin{figure}
\centering
\includegraphics[width=0.5\columnwidth]{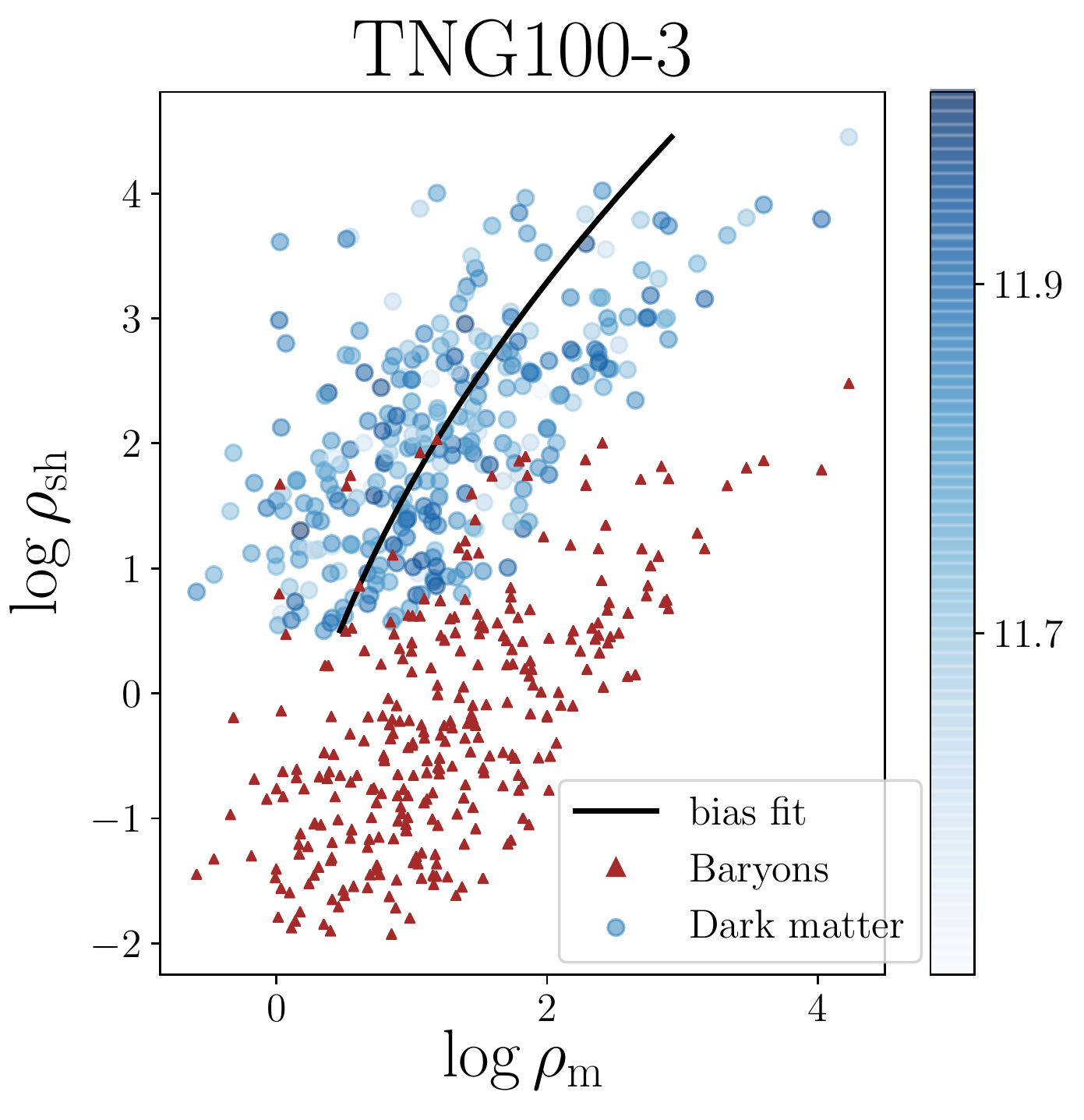} 
\caption{ 
(Colour online). Scatter plot of density-in-cells for the dark matter field ($\rho_{\rm m}$) versus density-in-cells for the subhaloes ($\rho_{\rm sh}$), extracted from the IllustrisTNG 100-3-Dark run at $z=0$, with a fixed cell radius of $R = 10$ Mpc $h^{-1}$. The fitting of the bias model, Eq. \ref{bias}, is shown. For comparison, the corresponding scatter plot for the density-in-cells of baryons only, within the same selected haloes, is shown (extracted from the corresponding IllustrisTNG 100-3 full run). Vertical colour bar indicates the values of $\log_{10} \langle M_s \rangle$ (the average mass in the respective cells). 
\label{Logdens}}
\end{figure}

\begin{figure}
\centering
\includegraphics[trim= 0in 3.0in 0in 0.1in,clip,width=1.0\linewidth]{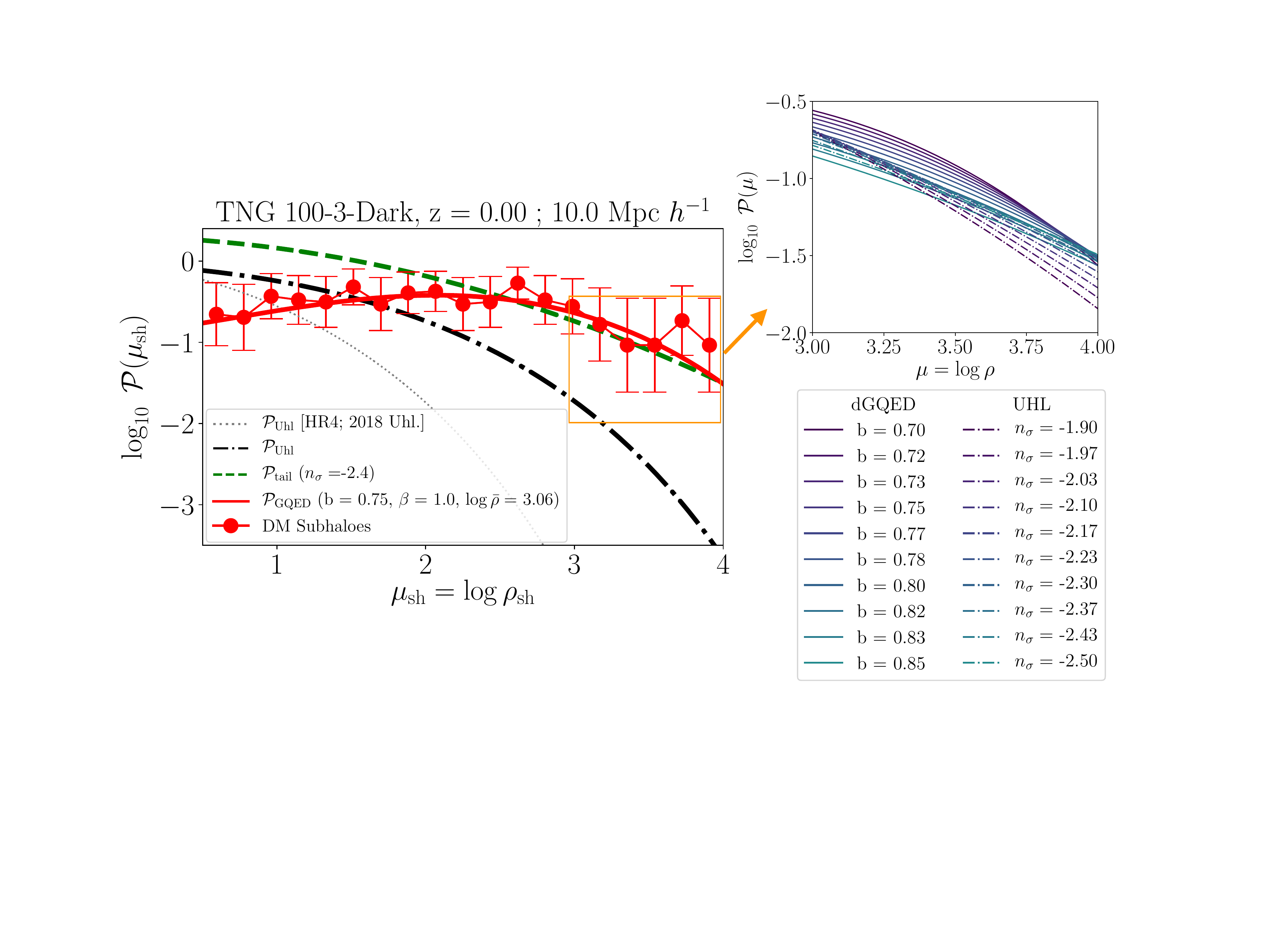}
\caption{
(Colour online). {\it Left panel}: Normalized histogram of the density-in-cells of dark matter subhaloes in the IllustrisTNG 100-3-Dark run at $z=0$ (error bars are given by the normalized bin counts, weighted by the nonlinear rms variance of the log-density of subhaloes). The PDF curves for the log-densities of subhaloes (in decimal logarithmic scale) are predictions from: {\it (i)} the UHL PDF (Eq. \ref{PDFSH}; dot-dashed line), based on the measured dark matter density field variance and best-fit bias parameters (see Fig. \ref{Logdens}); {\it (ii)} the high-density tail of the UHL PDF (Eq. \ref{TAIL}, normalized; dashed line); and {\it (iii)}  the dGQED PDF (Eq.  \ref{GQEDPDF}, normalized; continuous line). Also shown is the $\mathcal{P}_{\rm Uhl}$ curve for the Horizon Run 4 (thin dotted line) at the same $z$ and $R$ values \citep{Uhl18}. {\it Right panel}: Detail of the predictions of the PDF models in the high density region, as indicated by the box and arrow in the left panel. We show a comparison between the high density tail of the UHL PDF (dot-dashed lines), for a range of $n_{\sigma}$, and the dGQED PDF (continuous lines), for a range of $b$ (with $\log \bar{\rho} = 3.06, \beta =1.0$ fixed).
\label{Match}}
\end{figure}

Here we report the results of the density-in-cells statistics for the IllustrisTNG 100-3-Dark run at $z=0$. In Fig. \ref{Logdens}, we present the fitting of the bias model, Eq. \ref{bias}, over the  scatter plot of density-in-cells for the DM field  (obtained from the particle snapshot data) versus density-in-cells for subhaloes (obtained from the subhaloes catalogue). We also show, for comparison, the corresponding scatter plot for the density-in-cells of baryons only, within the same selected haloes (obtained from the IllsutrisTNG 100-3 full run at $z=0$). The bias best-fit parameters were: $b_0 = 0.297, b_1 = 0.312, b_2 = 0.062$. The measured nonlinear variances in the log-densities were: $\sigma_{\mu,{\rm m}}^2 = 0.669, \sigma_{\mu,{\rm sh}}^2 = 0.755$.

In Fig. \ref{Match} (left panel), we present the normalized histogram of the density-in-cells of DM subhaloes, representing the probability density function at each bin (the counting was normalized in order to give a unit integral over the range). Based on the bias best-fit parameters, the predicted PDF for the log-densities of subhaloes, $\mathcal{P}_{\rm Uhl}$ (Eq. \ref{PDFSH}; see also Eq. \ref{DMPDF}), was obtained.  We also plot the corresponding  high density tail of the PDF (Eq. \ref{TAIL}); in this case, a normalization of the curve was applied to achieve a good fit for the histogram at $\mu_{\rm sh} > 2.0$, with $n_{\sigma} = -2.4$. The dGQED PDF, $\mathcal{P}_{\rm GQED}$ (Eq. \ref{GQEDPDF}), was also plotted in Fig. \ref{Match} for comparison. In this case, we inverted Eq. \ref{sigmadGQED} for the expected $b$ parameter, using $\sigma_{\mu,{\rm sh}}^2 = 0.755$ , giving a good fit for $\log \bar{\rho} = 3.06$; a normalization of the curve was also applied.

As a reference, we also show in Fig. \ref{Match} (left panel) the approximate $\mathcal{P}_{\rm Uhl}$  curve for the Horizon Run 4 (HR4) simulation \citep{Kim15}, obtained in \cite{Uhl18}, also for $z=0$ and $R = 10$ Mpc $h^{-1}$. The computed $\mathcal{P}_{\rm Uhl}$ curves for the IllustrisTNG and  HR4 evidently differ, given the differences in those simulations, such as cosmological model, box size, range of resolved halo masses, etc. The HR4 is a $\Lambda$CDM cosmological N-body simulation, set with the WMAP-5 cosmology parameters, differing from the Planck-16 values used in the IllustrisTNG. Also, the HR4 is $3.15$ Gpc $h^{-1}$ box, much larger than the IllustrisTNG 100. The selected haloes in the HR4, used to validate the ULH PDF, have masses ranging from $2.7 \times  10^{11} ~{\rm M}_{\odot}h^{-1} $ to  $4.2 \times  10^{15} ~{\rm M}_{\odot} h^{-1} $, to be contrasted with our selection of subhalo masses, from $2.5 \times 10^8 ~{\rm M}_{\odot}h^{-1}$ to $\sim 10^{13} ~{\rm M}_{\odot}h^{-1}$. Also as a reference, for a fitting to the UHL PDF in the IllustrisTNG 100 at $R = 5$ Mpc $h^{-1}$ and $z = \{1, 2, 3, 4 ,5\}$, see \cite{Lei19}.

Our results indicate that the  (appropriately normalized) $\mathcal{P}_{\rm tail}$ (Eq. \ref{TAIL}) of the UHL PDF fits well the high density range of the extracted density-in-cells statistic, but not the saddle-point approximation $\mathcal{P}_{\rm Uhl}$ (Eq. \ref{PDFSH}), which has proven valid up to variances of the dark matter log-density of $\sigma_{\mu}^2 \sim 0.5$ \citep{Uhl16}. The variances in the log-densities obtained in our sampling ($\sigma_{\mu,{\rm m}}^2 = 0.669, \sigma_{\mu,{\rm sh}}^2 = 0.755$) are somewhat above that limit, which might explain the mismatch. Nevertheless, the  UHL PDF is compatible with the data for the limited intermediate range of $1 \lesssim \log \rho_{\rm sh} \lesssim 2$. The very low density range of the UHL PDF predicts a slight excess over the extracted data. Note that the high density tail required a value of $n_{\sigma} = -2.4$, which is at the limit to avoid the criticality of the decay-rate function (see Fig. 1 in \citealt{Uhl16}); nevertheless the fitting was adequate in that range. On the other hand, the (appropriately normalized) dGQED PDF fitted very well the entire range of extracted density-in-cells. 

We point out that, as the clustering parameter $b$ increases (for a fixed $\bar{\rho}$; cf. the right panel in Fig. \ref{DiC_PDF_GQED}), the dGQED PDF becomes flatter and skews in comparison to a Poissonian PDF. A qualitatively similar behaviour is seen in the UHL PDF, in terms of the variances of the density-in-cells statistics, as can be seen in Fig. A1 of \cite{Cod16}, in which the PDF deviates from a Gaussian (at low variances) to a very skewed distribution towards the high density range (at larger variances). This effect is also a function of time, as the initial PDF becomes gradually more skewed at lower redshifts, as voids increase in extent and density peaks increase in amplitude (clustering increases) via accretion of matter.

Interestingly, the high density tail of the UHL PDF approaches well the dGQED curve in that range. We investigated this proximity by plotting together both (normalized) PDFs in Fig. \ref{Match} (right panel), in the high density region, for a range of $n_{\sigma}$ in the case of the UHL PDF (Eq. \ref{TAIL}) and for a range of $b$ in the case of the dGQED PDF, fixing $\log \bar{\rho} = 3.06, \beta =1.0$  (Eq. \ref{GQEDPDF}). Clearly, the differences in the predictions can reach a low percentage or even subpercentage levels in that density range.

$~$

%%%%%%%%%%%%%%%%%%%%%%%%%%%%%%%%%%%%%%%%%%%%%%%%%%
% 4 - SUMMARY AND CONCLUSIONS
%%%%%%%%%%%%%%%%%%%%%%%%%%%%%%%%%%%%%%%%%%%%%%%%%%
\section{Summary and conclusions \label{DIS}}

We analyzed the compatibility of the one-point statistics of subhaloes in the IllustrisTNG simulations with predictions of four models: the GQED, the NBD, the dGQED and the UHL PDFs.  We extracted $136$ CiC samples from the IllustrisTNG 300-3, 100-3 and 100-1 (full and dark-only) runs, in a range of cell sizes and redshifts. For the density-in-cells extraction, we used the IllustrisTNG 100-3-Dark at $z=0$, at a fixed cell radius of $R = 10$ Mpc $h^{-1}$.

We found that both the full and dark-only runs follow similar GQED and NBD CiC PDF forms. The two simulation boxes cover similar ranges in CiC number counts.  Comparing our results in the literature, we found similar scaling and evolutionary trends in all samples, up to factors in the amplitude of parameter values in the case of observational data, possibly regulated by different magnitude cutoffs. Despite the similarity of the CiC PDFs, we found measurable differences between full and dark-only runs, leading to trends in the fitting parameters, which might be relevant for the understanding of bias in terms of gravithermodynamical predictions. For example, the clustering parameter $b$ in the full runs converged approximately to the dark-only runs at lower redshifts, but then $\bar{N}$ tended to become smaller relatively to the dark-only runs. This could be an indication that subhaloes in the full runs were merging inside common dark matter haloes more efficiently than in the dark-only runs.

The UHL PDF in the saddle-point approximation was compatible with an intermediate range of densities, with the (normalized) high density tail separately giving a good fit. The (normalized) dGQED PDF fitted very well the entire range of extracted data. Interestingly, we found that, after normalization, dGQED and UHL PDFs in the high density range approximated each other to subpercentage levels  for different parameters. This sector in the PDFs corresponding to rare events, namely, large density fluctuations, are important to constraint the dynamics and cosmology from the initial conditions in the density field to the final distribution \cite[e.g.][]{Cod16, Uhl18, Lei19, WKS20}.

Our work attempted to extend the scope of applications of the IllustrisTNG simulations into tests of gravithermodynamical theory under complex, multi-component physics, and to compare its performance against other predictions. For the most part, we found that the gravitational quasi-equilibrium thermodynamical assumptions still hold in the presence of baryonic physics, with residues increasing for the smallest cell size.  Given the open problem concerning the physical basis of the NBD  \citep{Sas96}, the meaning of adequate fittings to this PDF is unclear as to the level of its flexibility against variability of the data. The qualitative differences and similarities encoded in the analyzed one-point PDFs could enable an increased understanding of their common elements of validity, or assumptions to be discarded or modified. For instance, our results suggest that a connection between LDT and gravithermodynamics, in the case of high density events, could lead to novel insights. Such a development may be relevant for providing specific predictions for future large galaxy surveys.

%%%%%%%%%%%%%%%%%%%%%%%%%%%%%%%%%%%%%%%%%%%%%%%%%%
\section*{Acknowledgements}

CCD thanks dr. Hugo V. Capelato for his encouragement during the development of this project. CCD thanks dr. Cora Uhlemann helpful clarifications. CCD also thanks the referee for corrections and feedback, which greatly improved this work.

%%%%%%%%%%%%%%%%%%%%%%%%%%%%%%%%%%%%%%%%%%%%%%%%%%
\appendix

\section {Analysis of residues  \label{App}}

% trim 0 inches on the left, 1.3 inches on the bottom, 0 inches on the right, and 1.3 inches on the top. 
%
%\begin{figure}
%\centering
%\includegraphics[trim= 0in 3in 0in 1.5in,clip,width=1.0\linewidth]{FVN100.pdf} 

\begin{figure}
\centering
\includegraphics[trim= 0in 2in 0in 1.5in,clip,width=1.0\linewidth]{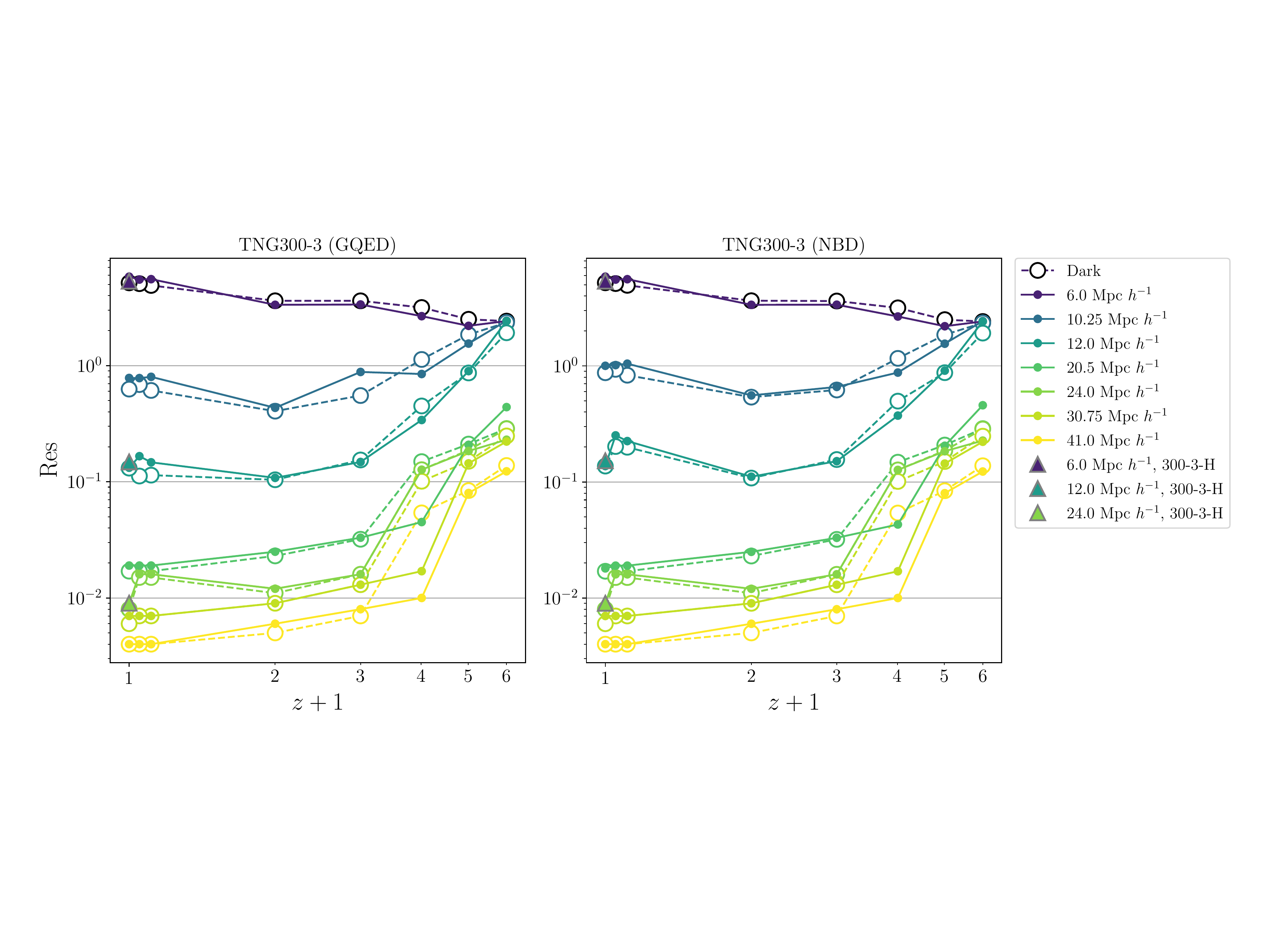} 
\caption{ (Colour online).  Residual 2-norms (in logarithmic scale) of the TNG300-3 CiC fittings as a function of redshift: GQED (left panel) and NBD (right panel). 
 \label{PARAMS-RES}}
\end{figure}

In this section we present a brief analysis of the residues in the CiC fits to the GQED and NBD models for TNG300-3 runs. The residues are shown in Fig. \ref{PARAMS-RES} as a function of redshift and cell size. As the counting cells are allowed to intersect, they form a statistical ensemble which is not entirely independent. Even if the cells were adjacent, objects belonging to nearby cells would be correlated. Hence, given the long range nature of the gravitational clustering, all cells are correlated in different degrees. In any case, the higher the number of cells in the ensemble used for the CiC computation, the better the statistics will be concerning the resulting form of $f_V(N)$. For the analysis of the goodness of the fitting models, we did not evaluate the $\chi^2$ estimates due to cell correlations; we use residual 2-norms of the fits, namely:

\begin{equation}
{\rm Res} = \sum_{N_0}^{N_{\rm max}} \left [ f_V(N)_{\rm sim} - f_V(N)_{\rm theo}\right ] ^2,
\end{equation}
\noindent where $N_0 \equiv (N = 0)$ and $N_{\rm max}$ is the largest number of galaxies in a cell.

We found that the residues (for both GQED and NBD models) were larger for smaller comoving cell sizes, specially for the smallest size. Except for the smallest cell, residues  tended to increase at higher redshifts. The TNG300-3-H case, in which twice of initial counting cells was used, showed somewhat smaller residues (relative to each cell size) than the TNG300-3 runs at $z=0$.  For brevity, we omitted similar figures for the TNG100-3 runs, which showed similar trends; we briefly note that the dark-only TNG100-3 runs showed a significantly larger residue than in its full counterpart for the smallest cell. Overall, both GQED and NBD models showed similar residues.

\section*{DATA AVAILABILITY STATEMENT}

The data underlying this article will be shared on reasonable request to the corresponding author.

%%%%%%%%%%%%%%%%%%%%%%%%%%%%%%%%%%%%%%%%%%%%%%%%%%%%%%%%%%%%%%%%%%%%%%%%%%%%%%%%%%%%%%%%%%
% BIBLIOGRAPHY
%%%%%%%%%%%%%%%%%%%%%%%%%%%%%%%%%%%%%%%%%%%%%%%%%%%%%%%%%%%%%%%%%%%%%%%%%%%%%%%%%%%%%%%%%%

\bibliographystyle{mnras}
\bibliography{CCDANTAS}

% Don't change these lines
\bsp	% typesetting comment
\label{lastpage}
\end{document}